\documentstyle[epsfig]{article}
\newcommand{\ba}{\begin{eqnarray}}
\newcommand{\ea}{\end{eqnarray}}
\def\ii{\'\i}
\begin{document}

\title{Randomness and Emerging Order in Nuclear Structure}
\author{R. Bijker$^1$ and A. Frank$^{1,2}$\\
\mbox{}\\
$^1$Instituto de Ciencias Nucleares, 
Universidad Nacional Aut\'onoma de M\'exico, \\
Apartado Postal 70-543, 04510 M\'exico, D.F., M\'exico \\
\mbox{}\\
$^2$Centro de Ciencias F{\'{\i}}sicas, 
Universidad Nacional Aut\'onoma de M\'exico, \\
Apartado Postal 139-B, Cuernavaca, Morelos, M\'exico}

\maketitle

\begin{abstract}
In order to investigate to what extent is the low-lying behavior of
even-even nuclei dependent on particular nucleon-nucleon interactions,
we consider systems of bosons where these interactions are taken as
gaussian random numbers with equal likelyhood of being attractive or
repulsive.  We find a statistical dominance of $L=0$ ground states and
other correlations, which we analyze in terms of a mean field approach.

\

Para investigar hasta que punto el comportamiento de los n\'ucleos
est\'a determinado por interacciones particulares entre los nucleones,
consideramos sistemas de bosones con interacciones gaussianas
aleatorias, donde dichas interacciones tienen la misma probabilidad de
ser atractivas o repulsivas.  Encontramos un dominio estad\ii stico de
estados base $L=0$ y otras correlaciones, que analizamos en t\'erminos
de un m\'etodo de campo medio.

\


\end{abstract}

\section{Introduction}

Whereas conventional wisdom ascribes the observed properties of
low-lying nuclear systems to a particular form of the nucleon-nucleon
interaction, recent investigations have found that some of these
features may represent a general and robust property of many-body systems 
\cite{JBD}-\cite{BFP2}.  Random matrices have been used in the past in
order to study generic spectral properties, but the emphasis was usually
centered on global characteristics, such as level-spacing distributions
in highly excited nuclei or the average properties of small metallic
particles or quantum dots \cite{Brody}. This approach had its
origin in the observation by E.P. Wigner that {\it `\dots the Hamiltonian 
which governs the behavior of a complicated system is a random symmetric
matrix, with no particular properties except for its symmetric
nature'}.  At high energy, nuclear dynamics is thus assumed to have lost
track of any correlations.  In the case of other systems, random
fluctuations in shapes and/or impurities can be reasonably modeled in
this form and their average properties reproduced.

A general conclusion of these studies is that the specific mechanism
(interactions) seem to be irrelevant in the determination of global
features.

However, the random-matrix analyses of nuclear distributions typically
involve states with the same quantum numbers, such as angular momentum,
parity or isospin, and little attention was given to correlations among
different symmetries. Although the so-called two-body random ensemble
(TBRE) was defined long time ago and some of its properties investigated
\cite{French,Bohigas}, again attention was focused mostly on its global 
features and not so much on low-lying patterns \cite{Brody}. 
This kind of ensemble is associated with
a random selection of the two-body matrix elements, while the $n$-body
Hamiltonian matrices are generated with the usual coupling procedures.
On the other hand, it is known that nuclei display very regular spectral
properties. For example, in Refs.~\cite{Casten,Zamfir} an analysis of
energy systematics of medium and heavy even-even nuclei suggests a
classification in terms of seniority, anharmonic vibrator and rotor
regions. Plots of the excitation energies of the yrast states with 
$L^P=4^+$ against $L^P=2^+$ show characteristic slopes for each region.
This and other correlations have been shown to be robust features of
low-energy nuclear behavior  which signal the emergence of order and
collectivity. In each case this behavior has been shown  to arise from 
particular nucleon-nucleon interactions, such as an attractive pairing 
force in semimagic nuclei and the strongly attractive neutron-proton 
quadrupole-quadrupole interaction for deformed nuclei. It came as
a surprise, therefore, that recent studies for even-even nuclei, in a
shell model with randomly distributed  two-body interactions (TBRE),
displayed a marked statistical preference for $L^P=0^+$ ground states,
energy gaps, and other indicators of ordered behavior \cite{JBD}. 
These results sparked a large number of other numerical and theoretical 
investigations, in both fermion \cite{Johnson,BFP1,JBDT,BFP2,MVZ,ZA} and 
boson systems \cite{BF1,BF2,BFP2,KZC,DK,BF3}, in an attempt to explain 
this remarkable and unexpected results. 

In this contribution we summarize some of the
most notable findings in the context of the Interacting Boson Model (IBM) 
and present a possible explanation in terms of a mean-field analysis. 
In the conclusions we indicate some of the many remaining open questions 
and future research.

\section{The interacting boson model}

\begin{figure}
\centerline{\hbox{\epsfig{figure=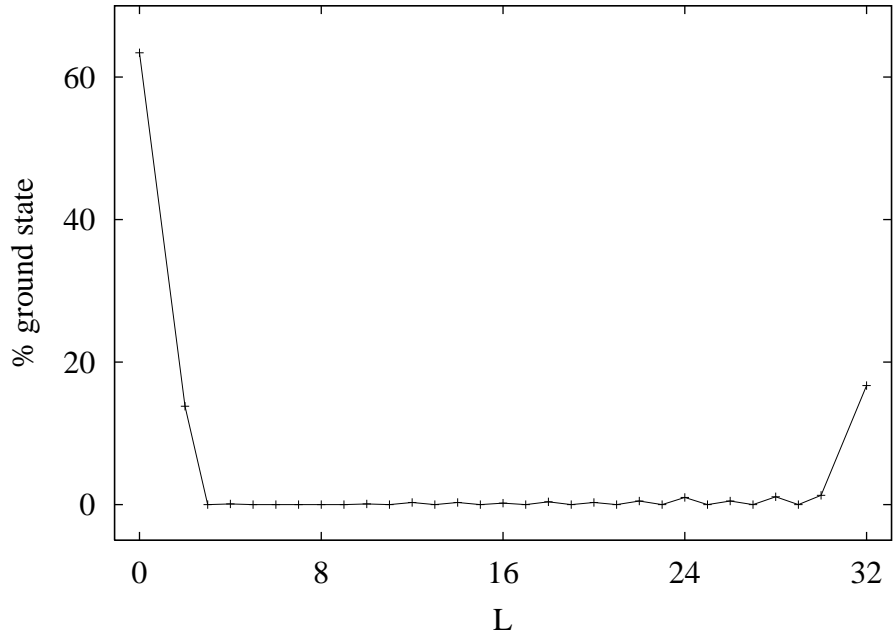} }}
\caption[]{Percentage of ground states with angular momentum $L$ 
in the IBM with random one- and two-body interactions obtained 
for $N=16$ and 1,000 runs.}
\label{ibmgs}
\end{figure}

\begin{figure}
\centerline{\hbox{\epsfig{figure=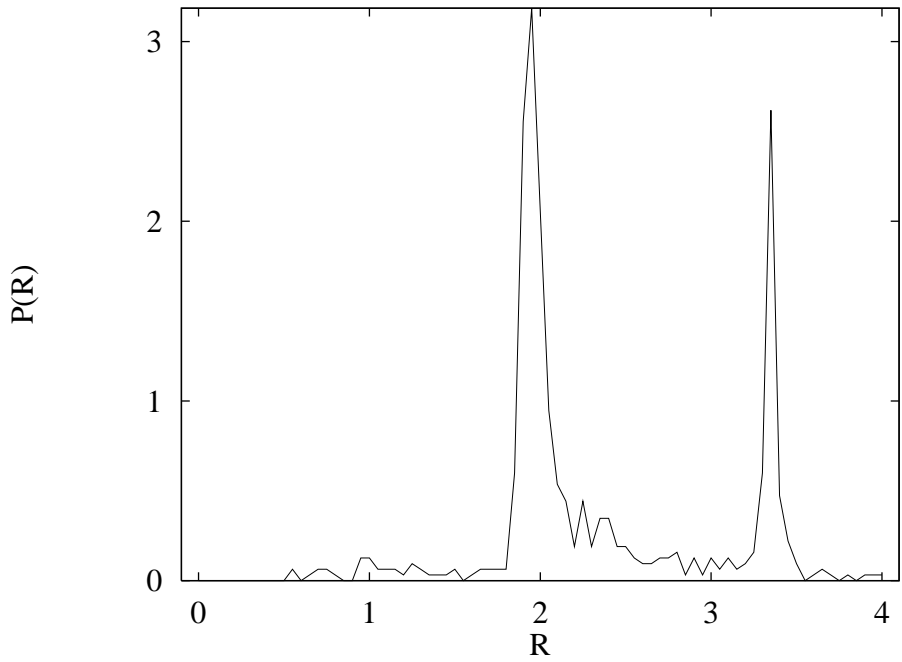} }}
\caption[]{Probability distribution $P(R)$ of the energy ratio $R$ 
of Eq.~(\protect\ref{ibmrat}) in the IBM with random one- and 
two-body interactions obtained for $N=16$ and 1,000 runs.}
\label{ibmpr}
\end{figure}

The IBM describes low-lying collective excitations in nuclei in terms 
of a system of $N$ interacting monopole and quadrupole bosons \cite{IBM}. 
An analysis of the IBM with random one- and two-body interactions gave 
rise to further statistical evidence of spontaneous emergence of order, 
manifested by a large percentage of ground states with $L^P=0^+$ and 
clear vibrational and rotational correlations \cite{BF1,BF2}. 
Fig.~\ref{ibmgs} shows 
that there is a predominance (63.4 $\%$) of $L^P=0^+$ ground states; 
in 13.8 $\%$ of the cases the ground state has $L^P=2^+$, and in 
16.7 $\%$ it has the maximum value of the angular momentum $L^P=32^+$. 
For the cases with a $L^P=0^+$ ground state, we show in Fig.~\ref{ibmpr} 
the probability distribution of the energy ratio 
\ba
R &=& \frac{E_{4^+_1}-E_{0^+_1}}{E_{2^+_1}-E_{0^+_1}} ~. 
\label{ibmrat}
\ea
There are two very pronounced peaks, right at the vibrational value 
of $R=2$ and at the rotational value of $R=10/3$, a clear indication of the 
occurrence of vibrational and rotational structure. This is confirmed 
by a simultaneous study of the quadrupole transitions between the levels 
\cite{BF1}. 

These are surprising results in the sense that, according to the conventional 
ideas in the field, the occurrence of $L=0$ ground states 
and the existence of vibrational and rotational bands are due to very 
specific forms of the interactions. The study of the 
IBM Hamiltonian with random one- and two-body interactions
seems to indicate that this may not be the entire story.  
However, the above results were obtained from numerical studies. 
It would be very interesting to gain a better understanding as to  
why this happens. What is the origin of the regular features which  
arise from random (both in sign and size) interactions? 
In this respect, there is a relevant quote by E.P. Wigner (as 
communicated to us by M. Moshinsky): {\it `I am happy to learn that the 
computer understands the problem, but I would like to understand it too'}.  

In the next sections, we report on an attempt in this direction by 
considering the vibron model, which has the same qualitative features 
as the IBM, namely vibrational and rotational spectra, but has a much 
simpler mathematical structure, since there are no coefficients of 
fractional parentage (cfp's). All many-body matrix elements can be 
derived in closed analytic form \cite{thebook,IL}. 

\section{The vibron model} 

The vibron model is an interacting boson model designed to describe the 
relative motion in two-body problems, e.g. diatomic molecules \cite{vibron}, 
nuclear clusters \cite{cluster} and mesons \cite{meson}. Its building 
blocks are a dipole boson $p^{\dagger}$ with $L^P=1^-$ and a 
scalar boson $s^{\dagger}$ with $L^P=0^+$. The total number of bosons $N$ 
is conserved by the vibron Hamiltonian. Here we only consider 
to one- and two-body interactions 
\ba
H &=& \frac{1}{N} \left[ H_1 + \frac{1}{N-1} H_2 \right] ~, 
\ea
where $H_1$ contains the boson energies 
\ba
H_1 &=& \epsilon_s \, s^{\dagger} \cdot \tilde{s}  
- \epsilon_p \, p^{\dagger} \cdot \tilde{p} ~, 
\ea
and $H_2$ all possible two-body interactions
\ba
H_2 &=& u_0 \, \frac{1}{2} \, 
(s^{\dagger} \times s^{\dagger})^{(0)} \cdot 
(\tilde{s} \times \tilde{s})^{(0)} 
+ u_1 \, (s^{\dagger} \times p^{\dagger})^{(1)} \cdot 
(\tilde{p} \times \tilde{s})^{(1)} 
\nonumber\\ 
&+& \sum_{\lambda=0,2} c_{\lambda} \, \frac{1}{2} \, 
(p^{\dagger} \times p^{\dagger})^{(\lambda)} \cdot  
(\tilde{p} \times \tilde{p})^{(\lambda)} 
\nonumber\\
&+& v_0 \, \frac{1}{2\sqrt{2}} \, \left[ 
  (p^{\dagger} \times p^{\dagger})^{(0)} \cdot 
  (\tilde{s} \times \tilde{s})^{(0)} 
+ H.c. \right] ~, 
\ea
and $\tilde{p}_m=(-1)^{1-m}p_{-m}$~. We have scaled $H_1$ by $N$ and 
$H_2$ by $N(N-1)$ in order to remove the $N$ dependence of the matrix 
elements. 

\begin{figure}
\centerline{\hbox{\epsfig{figure=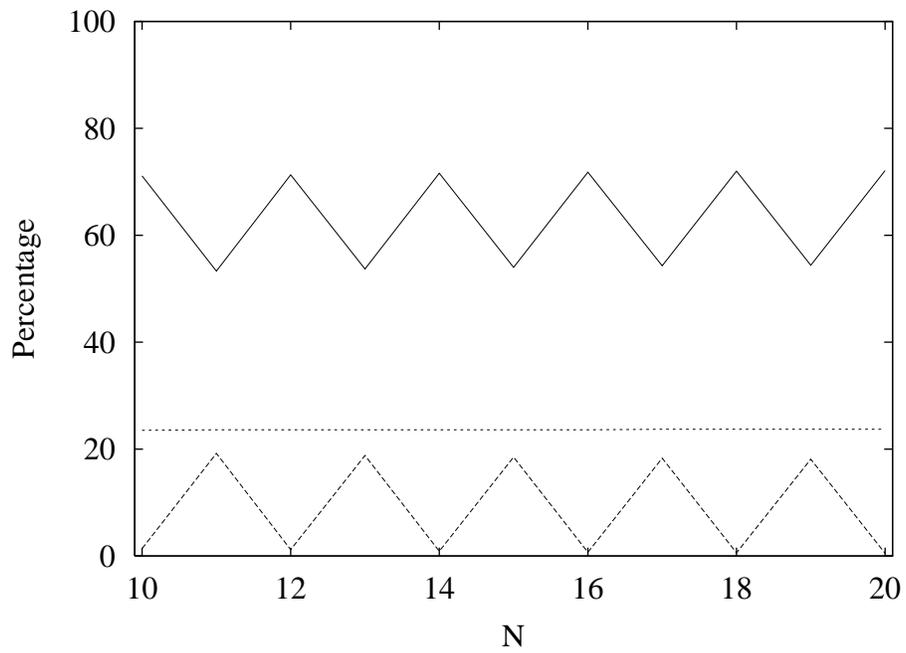} }}
\caption[]{Percentage of ground states with angular momentum $L=0$ 
(solid line), $L=1$ (dashed line) and $L=N$ (dotted line) in the vibron 
model with random one- and two-body interactions obtained for 
$10 \leq N \leq 20$ and 100,000 runs.}
\label{vibgs}
\end{figure}

We first carry out a numerical study of the vibron model using random 
one- and two-body interactions. The seven parameters of the Hamiltonian 
altogether denoted by 
\ba
(\vec{x}) &\equiv& (\epsilon_s, \epsilon_p, 
u_0, u_1, c_0, c_2, v_0) ~, 
\ea
are taken as independent random numbers on a Gaussian distribution 
\ba
P(x_i) &=& \mbox{e}^{-x_i^2/2\sigma^2}/\sigma\sqrt{2\pi} ~, 
\label{gauss}
\ea
with zero mean and width $\sigma$. In Fig.~\ref{vibgs} we show the 
percentages of $L=0$, $L=1$ and $L=N$ ground states as a function 
of the total number of vibrons $N$. Just as for the IBM, the vibron model 
shows a dominance of $L=0$ ground states. For even values of $N$ the 
ground state has $L=0$ in $\sim$ 71 $\%$ of the cases, and for odd 
values in $\sim$ 55 $\%$ of the cases. Similarly, the 
percentage of ground states with $L=1$ shows an oscillation between 
$\sim$ 1 $\%$ for even values of $N$ and $\sim$ 18 $\%$ for odd values. 
In $\sim$ 24 $\%$ of the cases the ground state has the maximum value 
of the angular momentum $L=N$. 

\begin{figure}
\centerline{\hbox{\epsfig{figure=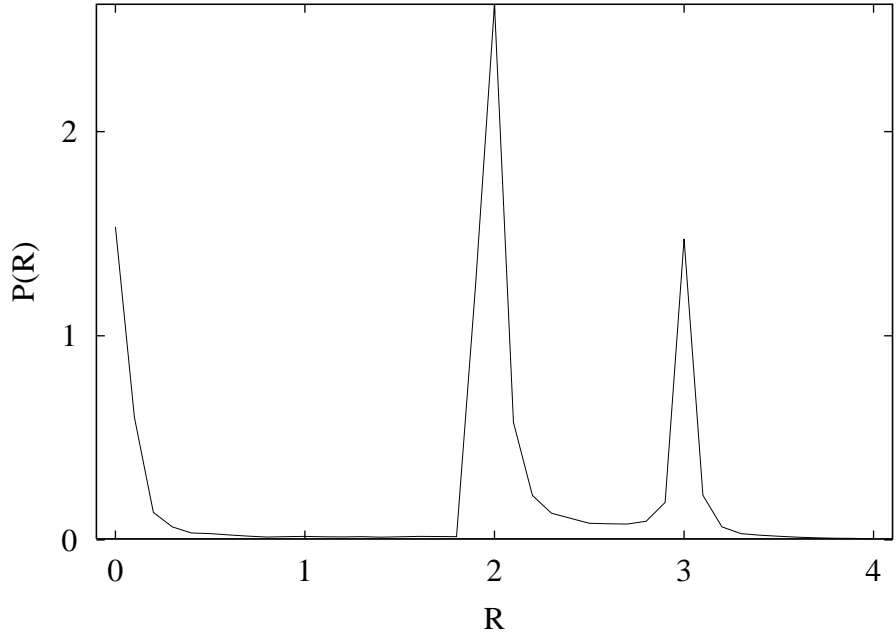} }}
\caption[]{Probability distribution $P(R)$ of the energy ratio $R$ 
of Eq.~(\protect\ref{vibrat}) in the vibron model with random one- and 
two-body interactions obtained for $N=20$ and 100,000 runs.}
\label{vibpr1}
\end{figure}

\begin{figure}
\centerline{\hbox{\epsfig{figure=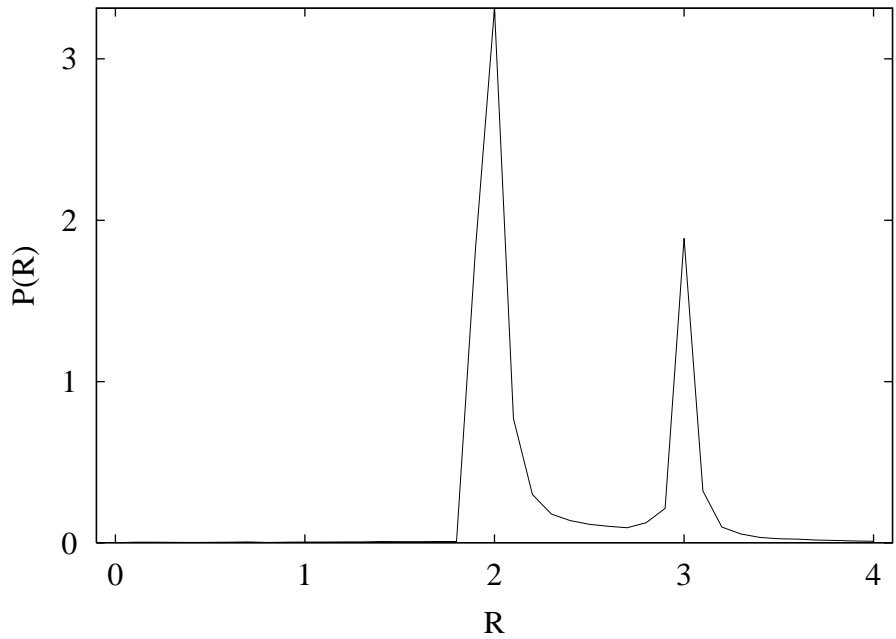} }}
\caption[]{As Fig.~\ref{vibpr1}, but for $N=21$.}
\label{vibpr2}
\end{figure}

For the cases with a $L^P=0^+$ ground state, we show in Figs.~\ref{vibpr1} 
and \ref{vibpr2} the probability distribution of the energy ratio 
\ba
R &=& \frac{E_{2^+_1}-E_{0^+_1}}{E_{1^-_1}-E_{0^+_1}} ~. 
\label{vibrat}
\ea
In the vibrational limit the spectrum is that of a three-dimensional 
harmonic oscillator with $R=2$, and in the rotational limit that of 
a three-dimensional Morse oscillator with $R=3$. Both for even and 
odd values of $N$ we see two clear peaks, one at the vibrational value 
of $R=2$ and one at the rotational value of $R=3$. Moreover, for even 
values of $N$ there is a peak at $R=0$, which is absent for odd values, 
as we shall prove below. 

This numerical study confirms that the vibron model exhibits the same 
regular features as does the IBM, although there are some differences 
as well. Now the question is, how can we understand these properties 
in an analytic and more intuitive way?  
In a recent study the tridiagonal form of the Hamiltonian matrix of 
the vibron model was used to establish a connection with random 
polynomials \cite{DK,BF3}. However, in general the Hamiltonian matrix 
is not of this form, and one has to look for alternative methods. 

In the next section we show that a mean-field study of the vibron model 
with random interactions can account for all features that were discussed 
above. 

\section{Mean-field analysis}

The connection between the vibron model, potential energy surfaces, 
equilibrium configurations, shapes, etc., can be studied by means of 
coherent states. The coherent state for the vibron model can be written 
as a condensate of a deformed boson which is a superposition of a scalar 
and a dipole boson \cite{onno}
\ba
\left| \, N,\alpha \, \right> \;=\; \frac{1}{\sqrt{N!}} \, 
\left( \sqrt{1-\alpha^2} \, s^{\dagger} + \alpha \, p_0^{\dagger} 
\right)^N \, \left| \, 0 \, \right> ~, 
\label{trial}
\ea
with $0 \leq \alpha \leq 1$. The potential energy surface is then given 
by the expectation value of the Hamiltonian in the coherent state 
\ba
E_N(\alpha) \;=\; \left< \, N,\alpha \, \right| \, H \, \left| \, 
N,\alpha \, \right> \;=\; a_4 \, \alpha^4 + a_2 \, \alpha^2 + a_0 ~, 
\label{surface}
\ea
where the coefficients $a_i$ are linear combinations of the parameters 
of the Hamiltonian 
\ba
a_4 &=& \vec{r} \cdot \vec{x} \;=\; \frac{1}{2} u_0 + u_1 
+ \frac{1}{6} c_0 + \frac{1}{3} c_2 + \frac{1}{\sqrt{6}} v_0 ~, 
\nonumber\\
a_2 &=& \vec{s} \cdot \vec{x} \;=\; -\epsilon_s + \epsilon_p 
- u_0 - u_1 - \frac{1}{\sqrt{6}} v_0 ~, 
\nonumber\\
a_0 &=& \epsilon_s + \frac{1}{2} u_0 ~. 
\label{coef}
\ea
For random interactions, we expect the trial wave function of 
Eq.~(\ref{trial}) and the energy surface of Eq.~(\ref{surface}) to 
provide information on the distribution of shapes that the model can 
acquire. The equilibrium configuration is characterized by the value of 
$\alpha=\alpha_0$ for which the energy surface $E_N(\alpha)$ has its 
minimum value. For a given Hamiltonian, the value of $\alpha_0$ depends 
on the coefficients $a_4$ and $a_2$. The distribution of shapes for 
an ensemble of Hamiltonians depends on the joint probability 
distribution of the coefficients $a_4$ and $a_2$ 
\ba
P(a_4,a_2) &=& \int \Pi_i \, dx_i \, P(x_i) \, 
\delta(a_4-\vec{r} \cdot \vec{x}) \, \delta(a_2-\vec{s} \cdot \vec{x})
\nonumber\\
&=& \frac{1}{2\pi \det M} \, \mbox{exp} \left[-\frac{1}{2} 
\left( \begin{array}{cc} a_4 & a_2 \end{array} \right) M^{-1} 
\left( \begin{array}{c} a_4 \\ a_2 \end{array} \right) \right] ~, 
\ea
where $P(x_i)$ is given by Eq.(\ref{gauss}) and 
\ba
M &=& \left( \begin{array}{ccc} 
\vec{r} \cdot \vec{r} &\hspace{0.5cm}& \vec{r} \cdot \vec{s} \\
\vec{r} \cdot \vec{s} & & \vec{s} \cdot \vec{s} \end{array} 
\right) ~. 
\ea
The vectors $\vec{r}$ and $\vec{s}$ are defined in Eq.~(\ref{coef}). 
In this approximation, we find that there are only three possible 
equilibrium configurations: 
\begin{itemize}

\item $\alpha_0=0$: $s$-boson condensate. This corresponds to a spherical 
shape whose probability can be obtained by integrating $P(a_4,a_2)$ over 
the appropriate range $S_1$ 
\ba
P_1 &=& \int_{S_1} \, da_4 \, da_2 \, P(a_4,a_2)  
\nonumber\\
&=& \frac{1}{4\pi} \left[ \pi + 2 \arctan \left( 
\frac{|\vec{s} \cdot \vec{s} + \vec{r} \cdot \vec{s}|}
{\sqrt{\det M}} \right) \right]  
\;=\; 0.396 ~. 
\ea
This configuration has spherical symmetry and hence can only have $L=0$. 

\item $0 < \alpha_0 < 1$: deformed condensate. The probability for 
the occurrence of a deformed shape can be calculated in a similar way 
and is given by 
\ba
P_2 &=& \frac{1}{2\pi} \arctan \left( 
\frac{2\sqrt{\det M}} 
{\vec{s} \cdot \vec{s} + 2\vec{r} \cdot \vec{s}} \right) 
\;=\; 0.216 ~.  
\ea
A deformed shape corresponds to a rotational band with angular momenta 
$L=0,1,\ldots,N$. 

\item $\alpha_0=1$. The probability for finding the third solution, 
a $p$-boson condensate, is given by 
\ba
P_3 \;=\; 1-P_1-P_2 \;=\; 0.388 ~. 
\nonumber
\ea
The angular momentum content of a $p$-boson condensate is 
$L=N,N-2,\ldots,1$ for $N$ odd or $L=N,N-2,\ldots,0$ for $N$ even. 

\end{itemize}
The probability that the ground state has $L=0$ can be estimated by 
assuming a complete decoupling of vibrations and rotations. For $N$ even 
this gives $\frac{1}{2}(1+P_1)$ or 69.8 $\%$, and for 
$N$ odd we have $P_1+\frac{1}{2}P_2$ or 50.4 $\%$, in good agreement 
with the numerical values. In Table~\ref{prob} we show the probabilities 
and corresponding percentages for the ground 
states with angular momentum $L=0$, $1$ and $N$. 

A better estimate can be obtained by evaluating the moment 
of inertia. We adopt the Thouless-Valatin prescription, which leads to 
the formula
\ba
{\cal I} \;=\; \frac{2N \alpha_0^2}{4(N-1)\left[ 
\frac{1}{2\sqrt{6}}v_0 (1-\alpha_0^2) - \frac{1}{6}(c_0-c_2) 
\alpha_0^2 \right]} ~.
\label{tv}
\ea
The moment of inertia thus depends in a complicated way on the parameters 
in the Hamiltonian, both explicitly as seen in the denominator of 
Eq.~(\ref{tv}) and implicitly through 
$\alpha_0$, which determines the equilibrium configuration. Formally, 
the probability to find a $L=0$ ground state is determined by the 
integral of the joint probability distribution $P(a_4,a_2,{\cal I})$ 
over the appropriate range. The results are summarized in 
Table~\ref{perc}. The spherical shape correponds to an energy ratio 
of $R=2$, and the deformed shape to $R=3$. This explains the vibrational 
and rotational peaks observed in Fig.~\ref{vibpr1} and \ref{vibpr2}. 
The occurrence of a peak at $R=0$ for even values of $N$ is related to 
the $p$-boson condensate solution, which has angular momenta $L=0,2,\ldots,N$. 
The first excited $L=1$ state belongs to a different band and has a higher 
excitation energy. For odd values of $N$ the $p$-boson condensate has no 
state with $L=0$, and hence the peak at $R=0$ is absent. 

\begin{figure}
\centerline{\hbox{\epsfig{figure=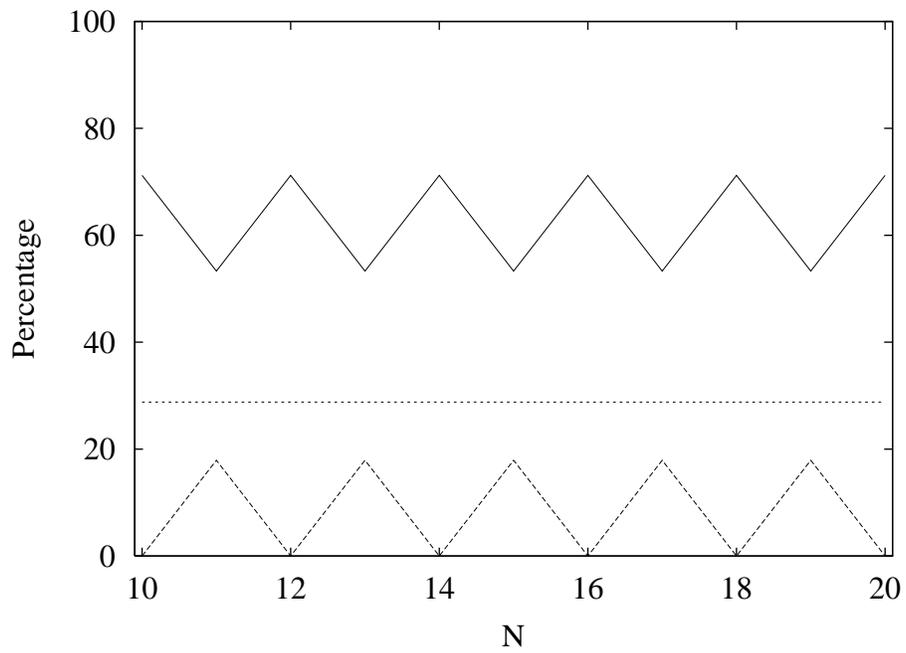} }}
\caption[]{Percentage of ground states with angular momentum $L=0$ 
(solid line), $L=1$ (dashed line) and $L=N$ (dotted line) in the vibron 
model with random one- and two-body interactions obtained in a mean-field 
analysis for $10 \leq N \leq 20$.} 
\label{vibmf}
\end{figure}

In Fig.~\ref{vibmf} we show the percentages of $L=0$, $L=1$ and $L=N$ 
ground states, as calculated in the mean-field analysis. A comparison with 
the exact results of Fig.~\ref{vibgs} shows that the mean-field results 
are in excellent agreement with the exact ones. The oscillations in the 
percentages of $L=0$ and $L=1$ ground states are entirely due to the 
angular momentum decomposition of the $p$-boson condensate, whereas the 
percentage of ground states with the maximum value of the angular momentum 
$L=N$ is a constant and does not depend on $N$. 

\section{Summary and conclusions}

In this contribution, we have studied the properties of low-lying
collective levels in the interacting boson model and the vibron model 
with random interactions. In particular, we discussed the dominance 
of $L=0$ ground states and the occurrence of vibrational and 
rotational band structures. In both models these features represent 
general and robust properties, and are a consequence of the many-body 
dynamics. 

We have studied the origin of these regular features in more detail 
in the context of the vibron model. In a mean-field analysis of the 
vibron model with random interactions it was found that we can account 
for all observed features, both qualitatively and quantitavely. 
The occurrence of $L=0$ ground states can be related to the different 
equilibrium configurations of the vibron model. 
There are three different equilibrium configurations: a spherical shape 
(40 $\%$), a deformed shape (20 $\%$) and a condensate of dipole bosons 
(40 $\%$). Since the spherical shape only has $L=0$, 
and the deformed shape and the $p$-boson condensate with $N$ even in 
about half the number of cases, one finds a $L=0$ ground state in 
approximately 70 $\%$ of the cases for $N$ even and 50 $\%$ for $N$ odd. 
The spherical shape gives rise to the occurrence of vibrational structure, 
and the deformed shape to rotational bands. 

The advantage of the mean-field method is that it can be applied directly 
to the IBM as well. The use of coherent states circumvents the use of 
coefficients of fractional parentage, and allows one to associate 
different regions of the parameter space with geometric shapes. 

There are many open questions in this area of research. Although the results 
in the case of the collective boson models can be explained by the allowed 
shape distributions, there is still no clear explanation for the dominance 
of $L=0$ ground states in fermion systems in the context of the shell 
model. For example, a recent analysis for single $j$-shell random Hamiltonians 
displays strong oscillations for the $L=0$ ground state percentage as 
a function of $j$ \cite{ZA}, reminiscent of the variations found for 
the IBM and the 
vibron model as a function of the boson number $N$ \cite{BF2}. The role of 
the approximate conservation of seniority should be further investigated. 
Another outstanding question is whether an appropriate truncation of the 
shell model space can lead to vibrational and rotational correlations of the 
kind observed in the IBM. Can shape be more precisely defined, for example, 
in a restricted fermion-paired Hilbert space? These and other studies 
are in progress. 

\section*{Acknowledgements}

We thank S. Pittel and  D. Rowe for interesting discussions. 
This work was supported in part by CONACyT under projects 
32416-E and 32397-E, and by DPAGA under project IN106400.

\clearpage

\begin{table}
\centering
\caption[]{Probabilities and percentages for ground states with 
angular momentum $L$ in the decoupling limit}
\label{prob}
\vspace{15pt}
\begin{tabular}{ccrcr}
\hline
& & & & \\
& \multicolumn{2}{c}{$N$ even} & \multicolumn{2}{c}{$N$ odd} \\
Ground State & Prob. & Perc. & Prob. & Perc. \\
& & & & \\
\hline
& & & & \\
$L=0$ & $\frac{1}{2}(1+P_1)$ & 69.8 $\,\%$  
& $P_1+\frac{1}{2}P_2$ & 50.4 $\,\%$ \\ 
& & & & \\
$L=1$ & $--$ & 0.0 $\,\%$ & $\frac{1}{2}(1-P_1-P_2)$ & 19.4 $\,\%$ \\ 
& & & & \\
$L=N$ & $\frac{1}{2}(1-P_1)$ & 30.2 $\,\%$  
& $\frac{1}{2}(1-P_1)$ & 30.2 $\,\%$ \\
& & & & \\
\hline
\end{tabular}
\end{table}

\begin{table}
\centering
\caption[]{Percentages of $L=0$ ground states, calculated in the 
decoupling limit (mf) and with the moment of inertia (${\cal I}$). 
The exact results are based on 100,000 runs with $N=20$ and $N=21$.}
\label{perc}
\vspace{15pt}
\begin{tabular}{cccccc}
\hline
& & & & & \\
& \multicolumn{2}{c}{$N$ even} & \multicolumn{2}{c}{$N$ odd} & \\
Shape & mf & ${\cal I}$ & mf & ${\cal I}$ & $R$ \\
& & & & & \\
\hline
& & & & & \\
$  \alpha_0=0$ & 39.6 $\,\%$ & 39.5 $\,\%$  
& 39.6 $\,\%$ & 39.5 $\,\%$ & 2 \\
& & & & & \\
$0<\alpha_0<1$ & 10.8 $\,\%$ & 13.8 $\,\%$ 
& 10.8 $\,\%$ & 13.8 $\,\%$ & 3 \\
& & & & & \\
$  \alpha_0=1$ & 19.4 $\,\%$ & 17.9 $\,\%$
&  0.0 $\,\%$ &  0.0 $\,\%$ & 0 \\
& & & & & \\
Total & 69.8 $\,\%$ & 71.2 $\,\%$ 
& 50.4 $\,\%$ & 53.3 $\,\%$ & \\
& & & & & \\
Exact & & 71.2 $\,\%$ & & 54.7 $\,\%$ & \\
& & & & & \\
\hline
\end{tabular}
\end{table}

\clearpage

\end{document}